\documentclass[twocolumn,english,aps,pra,superscriptaddress,showpacs,tightenlines]{revtex4-1}
\usepackage{amsmath}
\usepackage{graphicx}
\usepackage{amssymb}
\usepackage{CJK}
\usepackage{color}

\begin{document}

\begin{CJK*}{GBK}{song}

\title{resonator-assisted single-photon frequency convertion in a conventional waveguide with a giant V-type atom}
\author{Ge \surname{Sun} }
\affiliation{Key Laboratory of Low-Dimension Quantum Structures and Quantum Control of Ministry of Education, Key Laboratory for Matter Microstructure and Function of Hunan Province, Synergetic Innovation Center for Quantum Effects and Applications, Xiangjiang-Laboratory and Department of Physics, Hunan Normal University, Changsha 410081, China}
\affiliation{Institute of Interdisciplinary Studies, Hunan Normal University, Changsha, 410081, China}
\author{Hongzheng \surname{Wu} }
\affiliation{Key Laboratory of Low-Dimension Quantum Structures and Quantum Control of Ministry of Education, Key Laboratory for Matter Microstructure and Function of Hunan Province, Synergetic Innovation Center for Quantum Effects and Applications, Xiangjiang-Laboratory and Department of Physics, Hunan Normal University, Changsha 410081, China}
\affiliation{Institute of Interdisciplinary Studies, Hunan Normal University, Changsha, 410081, China}
\author{Jing \surname{Lu}}
\affiliation{Key Laboratory of Low-Dimension Quantum Structures and Quantum Control of Ministry of Education, Key Laboratory for Matter Microstructure and Function of Hunan Province, Synergetic Innovation Center for Quantum Effects and Applications, Xiangjiang-Laboratory and Department of Physics, Hunan Normal University, Changsha 410081, China}
\affiliation{Institute of Interdisciplinary Studies, Hunan Normal University, Changsha, 410081, China}
\author{Lan \surname{Zhou}}
\thanks{Corresponding author}
\email{zhoulan@hunnu.edu.cn}
\affiliation{Key Laboratory of Low-Dimension Quantum Structures and Quantum Control of Ministry of Education, Key Laboratory for Matter Microstructure and Function of Hunan Province, Synergetic Innovation Center for Quantum Effects and Applications, Xiangjiang-Laboratory and Department of Physics, Hunan Normal University, Changsha 410081, China}
\affiliation{Institute of Interdisciplinary Studies, Hunan Normal University, Changsha, 410081, China}

\begin{abstract}
We propose a scheme to achieve efficient frequency conversion for
a single photon propagating in a 1D conventional waveguide by exploiting the
quantum interference induced by the scale of a V-type giant atom (GA) characterized by the
distance between the two coupling points as well as single-photon transition
pathways originated from the coupling between the GA and the resonator.
The presence of photons in the resonator triggers the frequency conversion of photons. The scattering spectra
and the conversion contrast are studied in both the Markovian and the non-Markovian regimes.
The disappearance of frequency conversion is rooted in the complete suppression of the emission
from the excited state to either of lower states in the $n+1$ subspace where $n$ is the
photon number of the resonator, and the non-Markovicity-induced nonreciprocity is found
under specific conditions. Altering the photon number $n$ induces the non-reciprocal
transmission of single photons in the waveguide, hence, enhance the conversion probability.
\end{abstract}
\pacs{}
\maketitle

\end{CJK*}\narrowtext

\section{Introduction}

Quantum networks\cite{Kimble08} connect individual quantum systems via
quantum channels. Propagating photons are ideal flying qubits in quantum
channels for carry quantum information over long distances with almost
negligible decoherence. Precise control over single photons is required by
creasing large-scale, distributed quantum networks for technologies such as
computing, communication, sensing~\cite{RMP89QS}, and metrology~\cite%
{PRL96QM}. The lack of a direct interaction between photons at the
single-photon level in vacuum makes it necessary to rely on coupling with
quantum emitters (QEs). The coupling of QEs to propagating fields is
desirable in quantum network. However, the single-photon-single-QE coupling
is too weak in free space, the coupling can be substantially increased by
placing the QEs in waveguide structures~\cite{RMP89(17)021001}, which
confine photons to a one-dimensional (1D) environment. This kind of setup is
termed as waveguide-QED systems~\cite{RMP95(23)015002}.

To incorporate all the benefits needed, quantum systems in quantum networks
operate at dissimilar frequencies in order to perform each task at its
optimal frequency. It is desired to alter single photons from one frequency
to another frequency in order to implement hybrid quantum information
processing at different energy scales. Based on waveguide-QED systems,
frequency converter have been proposed for single photons by a point-like
three-level QE locally coupled to a Sagnac interferometer~\cite%
{QFC108PRLS,QFC108PRAs}, a coupled-resonator waveguide~\cite{QFC89PRAW}, a
semi-infinite 1D transmission line~\cite{QFC96PRAL}. With the great
technological progress to the chip scale, the waveguide-QED system has been
generalized to studies of a QE nonlocally coupled to photons in the
waveguide~\cite%
{ZLPRA85,LuOL49,LiAQT8(25),NM15NP1123,NM45OL3017,NM133PRL063603,GAE130PRL,GAE120PRL,GAE105PRA,GAE109PRA,GAE111PRA,GAE111PRAY,GAN103PRAY,GAN104PRAW}%
. The quantum emitter with dimension comparable to or larger than the
wavelength of its emitted photon is called giant atom (GA). A GA-waveguide
system can be realized by superconducting artificial QEs coupled to surface
acoustic waves~\cite{SAWSci346,SAWSNC8,SAWSPR124} or microwaves through
multiple coupling points through suitably meandering the transmission line
with wavelength-scale distance~\cite{GAWPRA103}. Frequency conversions for a
1D waveguide with a GA have been proposed in Ref.~\cite{QFC104PRAD,QFC3PRRD}%
. By employing the Markovian approximation, the conversion efficiency is at
most one-half in a conventional waveguide in Ref~\cite{QFC104PRAD}. Via the
chiral GA-waveguide couplings, the single-photon conversion efficiency is
enhanced to unity in Ref.~\cite{QFC3PRRD}. In this paper, we propose
single-photon frequency conversion for a single photon propagating in a 1D
\textit{conventional} waveguide scattered by a V-type GA coupled to a
single-mode resonator. For a conventional waveguide, the couplings strengths
of the GA to the right-going and left-going field modes are equal. The
frequency of the photon propagating in 1D waveguide is possible to be
converted up or down as long as photons are present in the resonator, and
the frequency shift is controlled by the number of photons in the resonator.
By utilizing a real-space scattering method~\cite{ShenPRL95,LanPRL101}, the
single-photon scattering spectra are obtained for a single photon incidence
from the waveguide and the number of excitations in the GA and the
resonator. We analyze the influences of phase delay between coupling points,
the phase difference between two GA-waveguide couplings and the number of
photons in the resonator on the scattering spectra in both the Markovian and
the non-Markovian regimes. It is found that the frequency conversion
vanishes when the emission from the excited state to either of lower states
is complete suppressed in the $n+1$ subspace where $n$ is the photon number
of the resonator. By adjusting the number of photons in the resonator, the
non-reciprocal transmission of single photons is present, and the conversion
probability for single photons can be achieved with unity efficiency.
\begin{figure}[tbp]
\includegraphics[width=8 cm,clip]{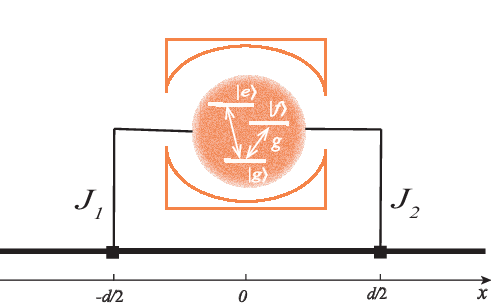}
\caption{Sketch of a $V$-type three-level atom coupled to a one-dimensional
waveguide. The transition $|g\rangle \leftrightarrow |e\rangle$ is coupled
to the waveguide at positions $x = \pm d/2$, while the transition $|g\rangle
\leftrightarrow |f\rangle$ is coupled to a single-mode cavity.}
\label{fig0}
\end{figure}


\section{\label{Sec:2}The System of a GA Coupled to A 1D Waveguide and A
Resonator}

We consider three energy levels $|g\rangle $,$|f\rangle $, $|e\rangle $ of a
single GA forming a $V$ configuration as illustrated in Fig.\ref{fig0}. The
ground state $|g\rangle $ is separated to state $|e\rangle $ ($|f\rangle $)
in frequency by $\omega _{e}(\omega _{f})$, and it can either transit to
state $|f\rangle $ by a single-mode resonator of frequency $\omega _{c}$
with $\hat{b}$ represented its annihilating operator, or transit to state $%
|e\rangle $ via absorbing a propagating photon in the waveguide at positions
$x_{1}=-d/2$, $x_{2}=d/2$. Denoting $\hat{a}_{L}(x)$ and $\hat{a}_{R}(x)$ as
the field operators of annihilating the right- and left-traveling photons at
position $x$ in waveguide, the full Hamiltonian of the system reads
\begin{eqnarray}
\hat{H} &=&\hat{H}_{0}+\left( g\hat{b}^{\dagger }\left\vert g\right\rangle
\left\langle f\right\vert +h.c.\right)  \label{Eq2-01} \\
&&+\sum_{j=1}^{2}\int dkJ_{j}\delta \left( x-x_{j}\right) e^{\left(
-1\right) ^{j}\mathrm{i}k_{0}\frac{d}{2}}\hat{a}_{L}^{\dagger }\left(
x\right) \left\vert g\right\rangle \left\langle e\right\vert +h.c.  \notag \\
&&+\sum_{j=1}^{2}\int dkJ_{j}\delta \left( x-x_{j}\right) e^{-\left(
-1\right) ^{j}\mathrm{i}k_{0}\frac{d}{2}}\hat{a}_{R}^{\dagger }\left(
x\right) \left\vert g\right\rangle \left\langle e\right\vert +h.c.  \notag
\end{eqnarray}%
with the free Hamiltonian%
\begin{eqnarray*}
H_{0} &=&\omega _{f}\left\vert f\right\rangle \left\langle f\right\vert
+\omega _{e}\left\vert e\right\rangle \left\langle e\right\vert +\omega _{c}%
\hat{b}^{\dagger }\hat{b} \\
&&+\int_{-\infty }^{\infty }dx\hat{a}_{R}^{\dagger }\left( x\right) \left(
\omega _{0}-\mathrm{i}v\frac{\partial }{\partial x}\right) \hat{a}_{R}\left(
x\right) \\
&&+\int_{-\infty }^{\infty }dx\hat{a}_{L}^{\dagger }\left( x\right) \left(
\omega _{0}+\mathrm{i}v\frac{\partial }{\partial x}\right) \hat{a}_{L}\left(
x\right) .
\end{eqnarray*}%
where coefficient $g$ and $J_{j}=\left\vert J_{j}\right\vert e^{i\varphi
_{j}}$ define the coupling strengths of the GA to the resonator and the
waveguide respectively, $\omega _{0}$ is the central frequency around which
a linear dispersion relation under consideration is given by $\omega \left(
k\right) =\omega _{0}+v\left( k-k_{0}\right) $, $v$ is the group velocity in
the vicinity of $\omega _{0}$. No direct coupling is assumed between the
resonator and waveguide.

We assume that the GA is initially in its ground state, the resonator
contains $n$ b-mode photons. A photon coming from either side of the
waveguide is scattered by the GA-resonator interaction. When $n=0$, the GA
absorbs the input photon and transit from $|g\rangle $ to its excited state $%
|e\rangle $, then back to its ground state by re-emitting the photon. In
this case, the system becomes a two-level GA interacting with the waveguide
since state $|f\rangle $ does not participates in the dynamic process, and
it is impossible for the frequency of the photon to be converted. As long as
$n\neq 0$, the transition $|e\rangle \leftrightarrow |f\rangle $ is allowed,
the quantum interference between different transitions of the GA leads to
behaviors different from $n=0$. The exchange of excitations between the GA
and the resonator mode raises (lowers) the state $|g\rangle $ ($|f\rangle $)
of the GA to the state $|f\rangle $ ($|g\rangle $) and lower (raise) the
number of photons in the resonator from $n$ ($n-1$) to $n-1$ ($n$). The
GA-resonator coupling changes the bare states of the GA and resonator to the
following dressed states
\begin{subequations}
\label{Eq2-02}
\begin{eqnarray}
\left\vert n_{+}\right\rangle &=&\sin \frac{\theta _{n}}{2}\left\vert
gn\right\rangle +\cos \frac{\theta _{n}}{2}\left\vert fn-1\right\rangle \\
\left\vert n_{-}\right\rangle &=&-\cos \frac{\theta _{n}}{2}\left\vert
gn\right\rangle +\sin \frac{\theta _{n}}{2}\left\vert fn-1\right\rangle
\end{eqnarray}%
with the corresponding eigenenergies $\lambda _{n\pm }=n\omega _{c}+\nu
_{\pm }^{n}$ and $\tan \theta _{n}=2\sqrt{n}g/\omega _{fc}$, where $\omega
_{fc}=\omega _{f}-\omega _{c}$ and
\end{subequations}
\begin{equation}
\nu _{\pm }^{n}=\frac{\omega _{fc}\pm \sqrt{\omega _{fc}^{2}+4n\left\vert
g\right\vert ^{2}}}{2}.  \label{Eq2-03}
\end{equation}%
When the GA is excited by an incoming photon in the waveguide exactly on
resonance with the $\left\vert e\right\rangle \leftrightarrow \left\vert
n_{-}\right\rangle $ transition, it will spontaneously decay to either $%
\left\vert n_{-}\right\rangle $ or $\left\vert n_{+}\right\rangle $ and emit
a photon, a frequency down-conversion is achieved when the GA decays to
state $\left\vert n_{+}\right\rangle $. Alternatively, frequency
up-conversion can be realized by initializing the GA in state $\left\vert
n_{+}\right\rangle $ and an incoming photon on resonance with the $|e\rangle
\leftrightarrow \left\vert n_{+}\right\rangle $ transition. Thence, the
presence of photons in the resonator triggers frequency conversion.

Since the total number of excitations in this system is conversed, states $%
\hat{a}_{R}^{\dagger }(x)|n_{\pm }0\rangle $, $\hat{a}_{L}^{\dagger
}(x)|n_{\pm }0\rangle $, $|en0\rangle $ form basis of the subspace with $n+1$
excitations, and they can expand the eigenstate in this subspace as%
\begin{equation}
\left\vert \psi _{n+1}\right\rangle =\sum_{P,\alpha }\int_{-\infty }^{\infty
}dxP_{\alpha }^{n}\left( x\right) \hat{a}_{p}^{\dagger }\left( x\right)
\left\vert n_{\alpha }0\right\rangle +u_{ne}\left\vert en0\right\rangle
\label{Eq2-04}
\end{equation}%
with $P=R,L$ and $\alpha =\pm $. Here, $R_{\alpha }^{n}\left( x\right) $ ($%
L_{\alpha }^{n}\left( x\right) $) is the single-photon wave function of a
right-moving (left-moving) photon in the waveguide at position $x$ and the
GA-resonator in state $\left\vert n_{\alpha }\right\rangle $, $u_{ne}$ is
the probability amplitude for all modes of the waveguide in vacuum states,
the resonator mode in the number state $n$, and the 3GA in its excited
state. The Schr\"{o}dinger eigen-equation yields the equations for the field
amplitudes
\begin{subequations}
\label{Eq2-05}
\begin{eqnarray}
ER_{+}^{n}\left( x\right) &=&\left( \omega _{+}^{n}-\mathrm{i}v\frac{%
\partial }{\partial x}\right) R_{+}^{n}\left( x\right) +V_{n+}^{R}u_{ne}, \\
EL_{+}^{n}\left( x\right) &=&\left( \omega _{+}^{n}+\mathrm{i}v\frac{%
\partial }{\partial x}\right) L_{+}^{n}\left( x\right) +V_{n+}^{L}u_{ne}, \\
ER_{-}^{n}\left( x\right) &=&\left( \omega _{-}^{n}-\mathrm{i}v\frac{%
\partial }{\partial x}\right) R_{-}^{n}\left( x\right) -V_{n-}^{R}u_{ne}, \\
EL_{-}^{n}\left( x\right) &=&\left( \omega _{-}^{n}+\mathrm{i}v\frac{%
\partial }{\partial x}\right) L_{-}^{n}\left( x\right) -V_{n-}^{L}u_{ne},
\end{eqnarray}
\end{subequations}
and the amplitude for the excited state of the GA
\begin{widetext}
\begin{eqnarray}
Eu_{ne} &=&\left( \omega _{e}+n\omega _{c}\right) u_{ne}  \notag \\
&&+J_{1+}^{\ast }\int dx\delta \left( x+\frac{d}{2}\right) \left[ e^{-%
\mathrm{i}k_{0}\frac{d}{2}}R_{+}^{n}\left( x\right) +e^{\mathrm{i}k_{0}\frac{%
d}{2}}L_{+}^{n}\left( x\right) \right]-J_{1-}^{\ast }\int dx\delta \left( x+\frac{d}{2}\right) \left[ e^{-%
\mathrm{i}k_{0}\frac{d}{2}}R_{-}^{n}\left( x\right) +e^{\mathrm{i}k_{0}\frac{%
d}{2}}L_{-}^{n}\left( x\right) \right]  \notag \\
&&+J_{2+}^{\ast }\int dx\delta \left( x-\frac{d}{2}\right) \left[ e^{\mathrm{%
i}k_{0}\frac{d}{2}}R_{+}^{n}\left( x\right) +e^{-\mathrm{i}k_{0}\frac{d}{2}%
}L_{+}^{n}\left( x\right) \right]-J_{2-}^{\ast }\int dx\delta \left( x-\frac{d}{2}\right) \left[ e^{\mathrm{%
i}k_{0}\frac{d}{2}}R_{-}^{n}\left( x\right) +e^{-\mathrm{i}k_{0}\frac{d}{2}%
}L_{-}^{n}\left( x\right) \right] ,  \notag \\
&&
\end{eqnarray}
\end{widetext}where we have defined
\begin{subequations}
\label{Eq2-06}
\begin{eqnarray}
\omega _{\pm }^{n} &=&\omega _{0}+n\omega _{c}+\nu _{\pm }^{n}, \\
J_{i+}^{n} &=&J_{i}\sin \frac{\theta _{n}}{2},J_{i-}^{n}=J_{i}\cos \frac{%
\theta _{n}}{2}, \\
V_{n\pm }^{R} &=&J_{1\pm }^{n}e^{\mathrm{i}k_{0}\frac{d}{2}}\delta \left( x+%
\frac{d}{2}\right) +J_{2\pm }^{n}e^{-\mathrm{i}k_{0}\frac{d}{2}}\delta
\left( x-\frac{d}{2}\right) ,  \notag \\
&& \\
V_{n\pm }^{L} &=&J_{1\pm }^{n}e^{-\mathrm{i}k_{0}\frac{d}{2}}\delta \left( x+%
\frac{d}{2}\right) +J_{2\pm }^{n}e^{\mathrm{i}k_{0}\frac{d}{2}}\delta \left(
x-\frac{d}{2}\right) .  \notag \\
&&
\end{eqnarray}

The wave function for the photon incident from the left-side takes the
following form:
\end{subequations}
\begin{subequations}
\begin{eqnarray}
R_{-}^{n}\left( x\right) &=&\left[ \Theta \left( -x-\frac{d}{2}\right)
+t_{-}^{n}\Theta \left( x-\frac{d}{2}\right) \right] e^{\mathrm{i}k_{n}x}
\notag \\
&&+A_{-}^{n}\Theta \left( x+\frac{d}{2}\right) \Theta \left( \frac{d}{2}%
-x\right) e^{\mathrm{i}k_{n}x}, \\
L_{-}^{n}\left( x\right) &=&\left[ r_{-}^{n}\Theta \left( -x-\frac{d}{2}%
\right) +B_{-}^{n}\Theta \left( \frac{d}{2}-\left\vert x\right\vert \right) %
\right] e^{-\mathrm{i}k_{n}x},  \notag \\
&& \\
R_{+}^{n}\left( x\right) &=&\left[ M^{n}\Theta \left( \frac{d}{2}-\left\vert
x\right\vert \right) +t_{+}^{n}\Theta \left( x-\frac{d}{2}\right) \right] e^{%
\mathrm{i}q_{n}x},  \notag \\
&& \\
L_{+}^{n}\left( x\right) &=&\left[ r_{+}^{n}\Theta \left( -x-\frac{d}{2}%
\right) +N^{n}\Theta \left( \frac{d}{2}-\left\vert x\right\vert \right) %
\right] e^{-\mathrm{i}q_{n}x},  \notag \\
&&
\end{eqnarray}
\end{subequations}%
where $\Theta \left( x\right) $ is the Heaviside step function, $t_{-}^{n}$
and $r_{-}^{n}$ are the single-photon transmission and reflection amplitudes
with wave vectors $k_{n}=\left( E-\omega _{-}^{n}\right) /v$ and $-k_{n}$
when the GA is in state $\left\vert n_{-}\right\rangle $, $t_{+}^{n}$ and $%
r_{+}^{n}$ are the single-photon conversion amplitudes with wave vectors $%
q_{n}=\left( E-\omega _{+}^{n}\right) /v$ and $-q_{n}$ when the GA is in
state $\left\vert n_{+}\right\rangle $. So the continuum of the waveguide
can be divided as a $n_{-}$-channel and a $n_{+}$-channel according to the
GA in state $\left\vert n_{-}\right\rangle $ or $\left\vert
n_{+}\right\rangle $. The transmittance $T_{-}^{n}=\left\vert
t_{-}^{n}\right\vert ^{2}$ and the reflectance $R_{-}^{n}=\left\vert
r_{-}^{n}\right\vert ^{2}$ describe the elastic scattering process which
preserves the original frequency of single photons. .The conversion
probability $T_{c}^{n}=|r_{+}^{n}|^{2}+|t_{+}^{n}|^{2}$ describes the
inelastic scattering process which changes the frequency of propagating
photons by the amount $|\lambda _{n+}-\lambda _{n-}|$. The scatter spectra
significantly depend on the spontaneous damping rates $\Gamma _{\pm }^{n}$
from the excited state $\left\vert e\right\rangle $ to state $\left\vert
n_{\pm }\right\rangle $ due to the coupling of the GA to the vacuum field,
and the nonlocal damping rates $\gamma _{\pm }^{n}$ from the coupling of the
GA through the vacuum field at different points
\begin{subequations}
\label{Eq2-07}
\begin{eqnarray}
\Gamma _{+}^{n} &=&\Gamma \sin ^{2}\frac{\theta _{n}}{2},\Gamma
_{-}^{n}=\Gamma \cos ^{2}\frac{\theta _{n}}{2} \\
\gamma _{+}^{n} &=&\frac{2\left\vert J_{1}J_{2}\right\vert }{v}\sin ^{2}%
\frac{\theta _{n}}{2}\cos \varphi _{J} \\
\gamma _{-}^{n} &=&\frac{2\left\vert J_{1}J_{2}\right\vert }{v}\cos ^{2}%
\frac{\theta _{n}}{2}\cos \varphi _{J}
\end{eqnarray}
\end{subequations}
with $\Gamma =\left( \left\vert J_{1}\right\vert ^{2}+\left\vert
J_{2}\right\vert ^{2}\right) /v$ and $\varphi _{J}=\varphi _{2}-\varphi _{1}$%
. The scattering amplitudes for a left-incident photon in the $n_{-}$%
-channel read
\begin{widetext}
\begin{subequations}
\label{Eq2-08}
\begin{eqnarray}
t_{-}^{n} &=&\frac{\Delta _{k}^{n}-2\frac{J_{1-}^{n\ast }J_{2-}^{n}}{v}\sin
\left( \Delta _{k}^{n}\tau +\phi _{-}^{n}\right) +\mathrm{i}\Gamma _{+}^{n}+%
\mathrm{i}\gamma _{+}^{n}e^{\mathrm{i}\Delta _{k}^{n}\tau }e^{\mathrm{i}\phi
_{+}^{n}}}{\Delta _{k}^{n}+\mathrm{i}\left( \Gamma _{+}^{n}+\gamma
_{+}^{n}e^{\mathrm{i}\Delta _{k}^{n}\tau }e^{\mathrm{i}\phi _{+}^{n}}\right)
+\mathrm{i}\left( \Gamma _{-}^{n}+\gamma _{-}^{n}e^{\mathrm{i}\Delta
_{k}^{n}\tau }e^{\mathrm{i}\phi _{-}^{n}}\right) }, \\
r_{-}^{n} &=&-\frac{\mathrm{i}\gamma _{-}+\frac{\mathrm{i}}{v}\left[
\left\vert J_{1-}^{n}\right\vert ^{2}e^{-\mathrm{i}\left( \Delta
_{k}^{n}\tau +\phi _{-}^{n}\right) }+\left\vert J_{2-}^{n}\right\vert ^{2}e^{%
\mathrm{i}\left( \Delta _{k}^{n}\tau +\phi _{-}^{n}\right) }\right] }{\Delta
_{k}^{n}+\mathrm{i}\left( \Gamma _{+}^{n}+\gamma _{+}^{n}e^{\mathrm{i}\Delta
_{n}\tau }e^{\mathrm{i}\phi _{+}^{n}}\right) +\mathrm{i}\left( \Gamma
_{-}^{n}+\gamma _{-}^{n}e^{\mathrm{i}\Delta _{k}^{n}\tau }e^{\mathrm{i}\phi
_{-}^{n}}\right) }, \\
r_{+}^{n} &=&\frac{\frac{\mathrm{i}e^{-\mathrm{i}\phi _{n}}}{v}\left[
J_{1+}^{n}e^{-\mathrm{i}\left( \Delta _{k}^{n}\tau +\phi _{n+}\right)
}+J_{2+}^{n}\right] \left[ J_{1-}^{n\ast }+J_{2-}^{n\ast }e^{\mathrm{i}%
\left( \Delta _{k}^{n}\tau +\phi _{-}^{n}\right) }\right] }{\Delta _{k}^{n}+%
\mathrm{i}\left( \Gamma _{+}^{n}+\gamma _{+}^{n}e^{\mathrm{i}\Delta
_{k}^{n}\tau }e^{\mathrm{i}\phi _{+}^{n}}\right) +\mathrm{i}\left( \Gamma
_{-}^{n}+\gamma _{-}^{n}e^{\mathrm{i}\Delta _{k}^{n}\tau }e^{\mathrm{i}\phi
_{-}^{n}}\right) }, \\
t_{+}^{n} &=&\frac{\frac{\mathrm{i}e^{-\mathrm{i}\phi _{n}}}{v}\left[
J_{1+}^{n}+J_{2+}^{n}e^{-\mathrm{i}\left( \Delta _{k}^{n}\tau +\phi
_{+}^{n}\right) }\right] \left[ J_{1-}^{n\ast }+J_{2-}^{n\ast }e^{\mathrm{i}%
\left( \Delta _{k}^{n}\tau +\phi _{-}^{n}\right) }\right] }{\Delta _{k}^{n}+%
\mathrm{i}\left( \Gamma _{+}^{n}+\gamma _{+}^{n}e^{\mathrm{i}\Delta
_{k}^{n}\tau }e^{\mathrm{i}\phi _{+}^{n}}\right) +\mathrm{i}\left( \Gamma
_{-}^{n}+\gamma _{-}^{n}e^{\mathrm{i}\Delta _{k}^{n}\tau }e^{\mathrm{i}\phi
_{-}^{n}}\right) },
\end{eqnarray}
\end{subequations}
\end{widetext}where $\Delta _{k}^{n}=\omega _{0}+vk_{n}-\left( \omega
_{e}-\nu _{-}^{n}\right) $ is the detuning between the input photon and the $%
\left\vert e\right\rangle \leftrightarrow \left\vert n_{-}\right\rangle $
transition, the accumulated phases
\begin{equation}
\phi _{\pm }^{n}=\left( \omega _{e}-\upsilon _{n\pm }\right) \tau ,\phi
_{n}=\left( \upsilon _{n+}-\upsilon _{n-}\right) \tau ,  \label{Eq2-09}
\end{equation}%
and $\tau =d/v$ is the delay time that photons travel along the distance
between two connection points.

For the photon incident from the far right of GA initial in the $n_{-}$%
-channel, scattering amplitudes read
\begin{widetext}
\begin{subequations}
\label{Eq2-10}
\begin{eqnarray}
\tilde{t}_{-}^{n} &=&\frac{\Delta _{k}^{n}-2\frac{J_{1-}^{n}J_{2-}^{n\ast }}{%
v}\sin \left( \Delta _{k}^{n}\tau +\phi _{-}^{n}\right) +\mathrm{i}\Gamma
_{+}^{n}+\mathrm{i}\gamma _{+}^{n}e^{\mathrm{i}\Delta _{k}^{n}\tau }e^{%
\mathrm{i}\phi _{+}^{n}}}{\Delta _{k}^{n}+\mathrm{i}\left( \Gamma
_{+}^{n}+\gamma _{+}^{n}e^{\mathrm{i}\Delta _{k}^{n}\tau }e^{\mathrm{i}\phi
_{+}^{n}}\right) +\mathrm{i}\left( \Gamma _{-}^{n}+\gamma _{-}^{n}e^{\mathrm{%
i}\Delta _{k}^{n}\tau }e^{\mathrm{i}\phi _{-}^{n}}\right) }, \\
\tilde{r}_{-}^{n} &=&-\frac{\mathrm{i}\gamma _{-}^{n}+\mathrm{i}\frac{%
\left\vert J_{1-}^{n}\right\vert ^{2}}{v}e^{\mathrm{i}\left( \Delta
_{k}^{n}\tau +\phi _{-}^{n}\right) }+\mathrm{i}\frac{\left\vert
J_{2-}^{n}\right\vert ^{2}}{v}e^{-\mathrm{i}\left( \Delta _{k}^{n}\tau +\phi
_{-}^{n}\right) }}{\Delta _{k}^{n}+\mathrm{i}\left( \Gamma _{+}^{n}+\gamma
_{+}^{n}e^{\mathrm{i}\Delta _{k}^{n}\tau }e^{\mathrm{i}\phi _{+}^{n}}\right)
+\mathrm{i}\left( \Gamma _{-}^{n}+\gamma _{-}^{n}e^{\mathrm{i}\Delta
_{k}^{n}\tau }e^{\mathrm{i}\phi _{-}^{n}}\right) }, \\
\tilde{r}_{+}^{n} &=&\frac{\frac{\mathrm{i}e^{\mathrm{i}\phi _{n}}}{v}\left(
J_{1+}^{n}e^{\mathrm{i}\Delta _{k}^{n}\tau }e^{\mathrm{i}\phi
_{+}^{n}}+J_{2+}^{n}\right) \left( J_{1-}^{n\ast }+J_{2-}^{n\ast }e^{-%
\mathrm{i}\Delta _{k}^{n}\tau }e^{-\mathrm{i}\phi _{-}^{n}}\right) }{\Delta
_{k}^{n}+\mathrm{i}\left( \Gamma _{+}^{n}+\gamma _{+}^{n}e^{\mathrm{i}\Delta
_{k}^{n}\tau }e^{\mathrm{i}\phi _{+}^{n}}\right) +\mathrm{i}\left( \Gamma
_{-}^{n}+\gamma _{-}^{n}e^{\mathrm{i}\Delta _{k}^{n}\tau }e^{\mathrm{i}\phi
_{-}^{n}}\right) }, \\
\tilde{t}_{+}^{n} &=&\frac{\frac{\mathrm{i}e^{\mathrm{i}\phi _{n}}}{v}\left(
J_{1+}^{n}+J_{2+}^{n}e^{\mathrm{i}\Delta _{k}^{n}\tau }e^{\mathrm{i}\phi
_{+}^{n}}\right) \left( J_{1-}^{n\ast }+J_{2-}^{n\ast }e^{-\mathrm{i}\Delta
_{k}^{n}\tau }e^{-\mathrm{i}\phi _{-}^{n}}\right) }{\Delta _{k}^{n}+\mathrm{i%
}\left( \Gamma _{+}^{n}+\gamma _{+}^{n}e^{\mathrm{i}\Delta _{k}^{n}\tau }e^{%
\mathrm{i}\phi _{+}^{n}}\right) +\mathrm{i}\left( \Gamma _{-}^{n}+\gamma
_{-}^{n}e^{\mathrm{i}\Delta _{k}^{n}\tau }e^{\mathrm{i}\phi _{-}^{n}}\right)
}.
\end{eqnarray}
\end{subequations}
\end{widetext}Similarly, the transmittance $\tilde{T}_{-}^{n}=\left\vert
\tilde{t}_{-}^{n}\right\vert ^{2}$ and the reflectance $\tilde{R}%
_{-}^{n}=\left\vert \tilde{r}_{-}^{n}\right\vert ^{2}$ describe the elastic
scattering process in the incident channel, which can be called $n_{-}$%
-channel since the GA is initially in $\left\vert n_{-}\right\rangle $. For
the GA ending in state $\left\vert n_{+}\right\rangle $, the photon is
scattered inelastically and propagates in the $n_{+}$-channel with the
conversion probability $\tilde{T}_{c}^{n}=|\tilde{r}_{+}^{n}|^{2}+|\tilde{t}%
_{+}^{n}|^{2}$. We will only study the transport properties for single
photons incidence from the $n_{-}$-channel since the discussion is similar
for photons incidence from the $n_{+}$-channel. The interaction with the
vacuum fluctuations of the waveguide generates the Lamb shift
\begin{equation}
\delta _{k}^{n}=\gamma _{+}^{n}\sin \left( \Delta _{k}^{n}\tau +\phi
_{+}^{n}\right) +\gamma _{-}^{n}\sin \left( \Delta _{k}^{n}\tau +\phi
_{-}^{n}\right)  \label{Eq2-11}
\end{equation}%
and the interference effects of reemitted photons from different coupling
points modified decay rates into the effective decay rate%
\begin{equation}
\tilde{\Gamma}_{k}^{n}=\sum_{\alpha =\pm }\left[ \Gamma _{\alpha
}^{n}+\gamma _{\alpha }^{n}\cos \left( \Delta _{k}^{n}\tau +\phi _{\alpha
}^{n}\right) \right] ,  \label{Eq2-12}
\end{equation}%
\begin{figure}[tbp]
\includegraphics[width=8 cm,clip]{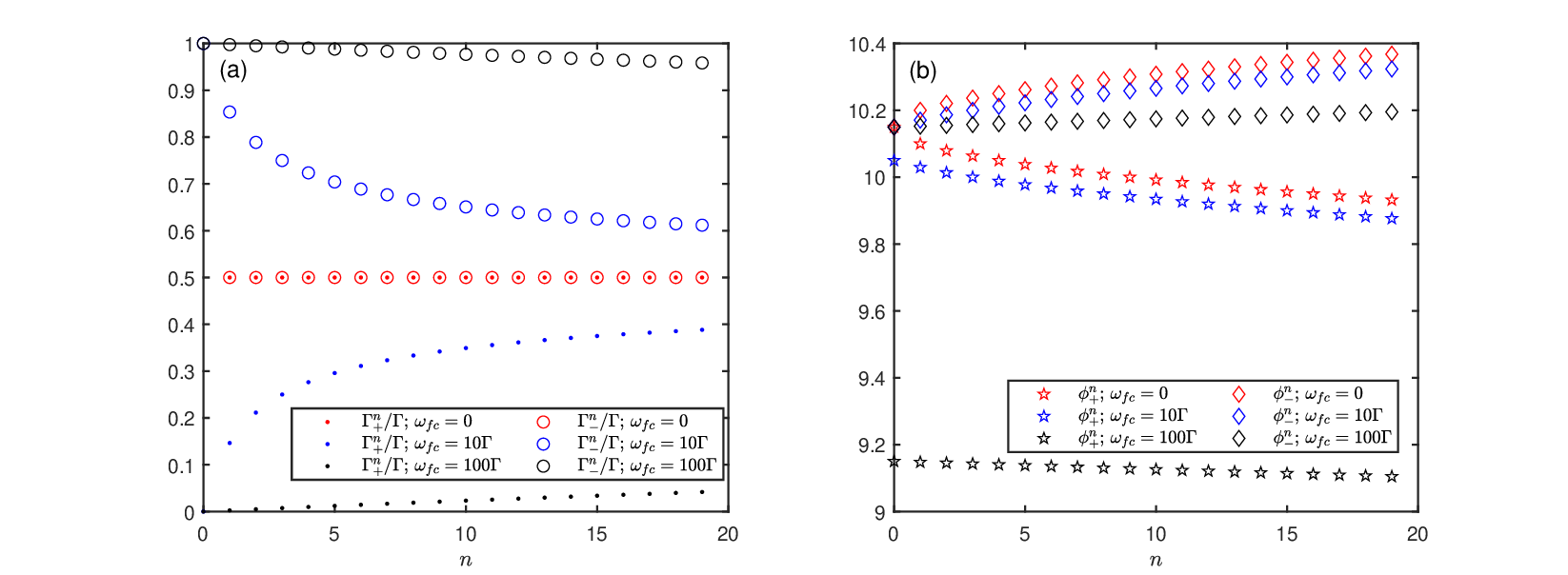}
\caption{(a) The spontaneous damping rates $\Gamma _{\pm }^{n}$ and (b) the
accumulated phases $\protect\phi _{\pm }^{n}$ versus photon number $n$ with $%
g=5\Gamma ,\protect\omega _{e}=1015\Gamma $, and $\Gamma \protect\tau =0.01$%
. }
\label{fig1}
\end{figure}
\begin{figure*}[tbp]
\includegraphics[width=0.9 \textwidth]{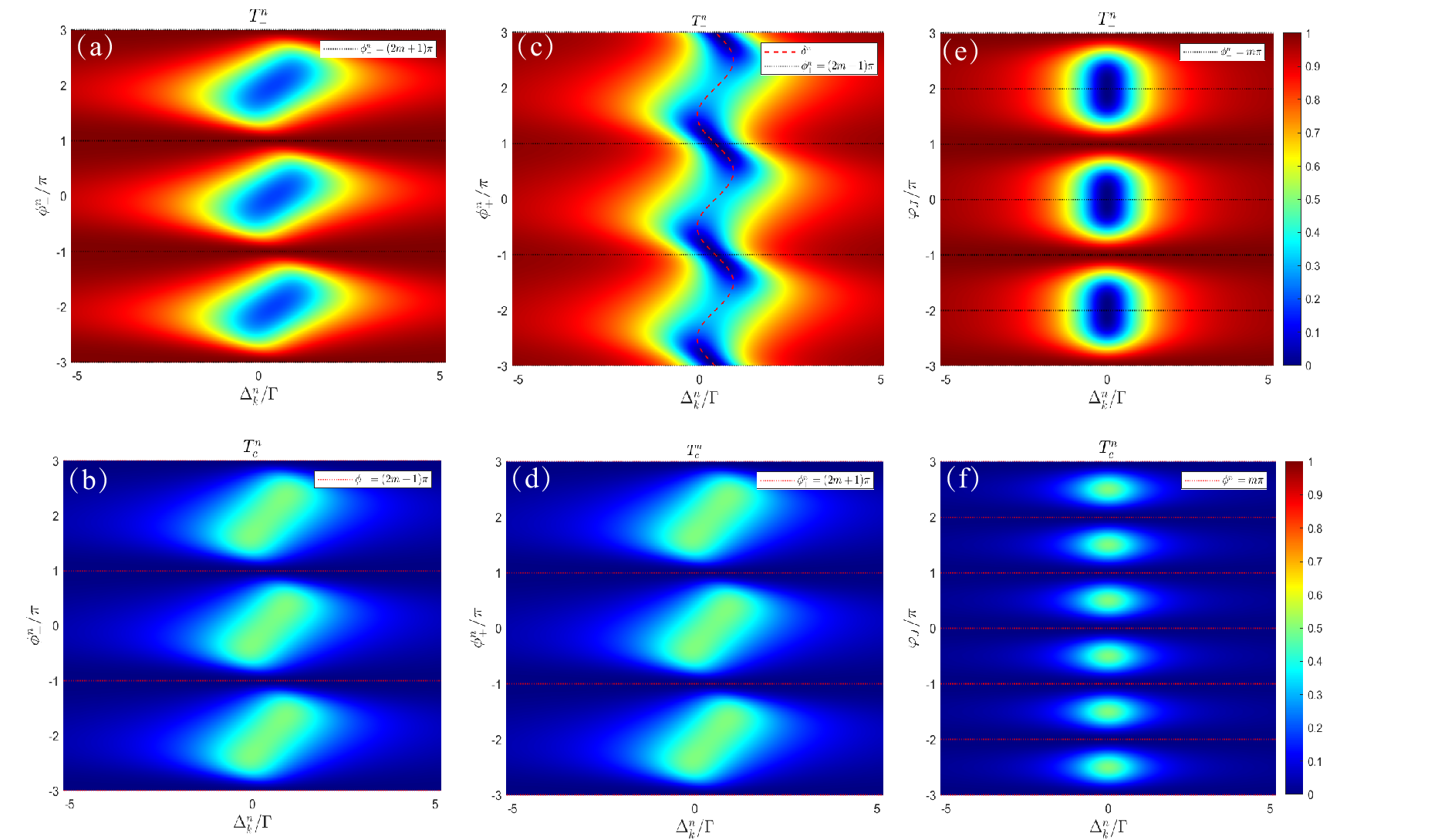}
\caption{(a,c,e) The transmittance $T_{-}^{\,n}$ and (b,d,f) the conversion
probability $T_{c}^{\,n}$ verse the scaled detuning $\Delta _{k}^{\,n}$ and
the scaled phase (a,b) $\protect\phi _{-}^{\,n}/\protect\pi $, (c,d) $%
\protect\phi _{+}^{\,n}/\protect\pi $, (e,f) $\protect\varphi _{J}/\protect%
\pi $ when $n=1$, $|J_{1}|=|J_{2}|$ and $\protect\omega _{fc}=0$. Other
parameters are setting as follow: (a,b) $\protect\varphi _{J}=0$ and $%
\protect\phi _{+}^{\,n}=\protect\pi /3$; (c,d) $\protect\varphi _{J}=0$ and $%
\protect\phi _{-}^{\,n}=\protect\pi /3$; (e,f) $\protect\phi _{-}^{\,n}=0$
and $\protect\phi _{+}^{\,n}=\protect\pi $. }
\label{fig2}
\end{figure*}
which change periodically with $\Delta _{k}^{n}$. It can be found from Eqs.(%
\ref{Eq2-08}) and (\ref{Eq2-10}) that the reflectance $R_{-}^{n}=\tilde{R}%
_{-}^{n}$, and the system presents reciprocity in transmission and
conversion upon reversing the propagation of an incident photon when $%
\varphi _{J}=n\pi $. To quantitatively describe the nonreciprocity, we
define the transmission contrast $I_{1}^{n}=T_{-}^{n}-\tilde{T}_{-}^{n}$ and
the conversion contrast $I_{2}^{n}=T_{c}^{n}-\tilde{T}_{c}^{n}$. The energy
conservation in the system indicates $I_{1}^{n}=-I_{2}^{n}$. The reciprocal
scattering corresponds to $I_{1}^{n}=I_{2}^{n}=0$. As the scattering
coefficients can not be smaller than zero, $|I_{1}^{n}|=|I_{2}^{n}|=1$
depicts the optimal nonreciprocal scattering corresponding to the unity
efficiency of the frequency conversion.
\begin{figure*}[tbp]
\includegraphics[width=0.9 \textwidth]{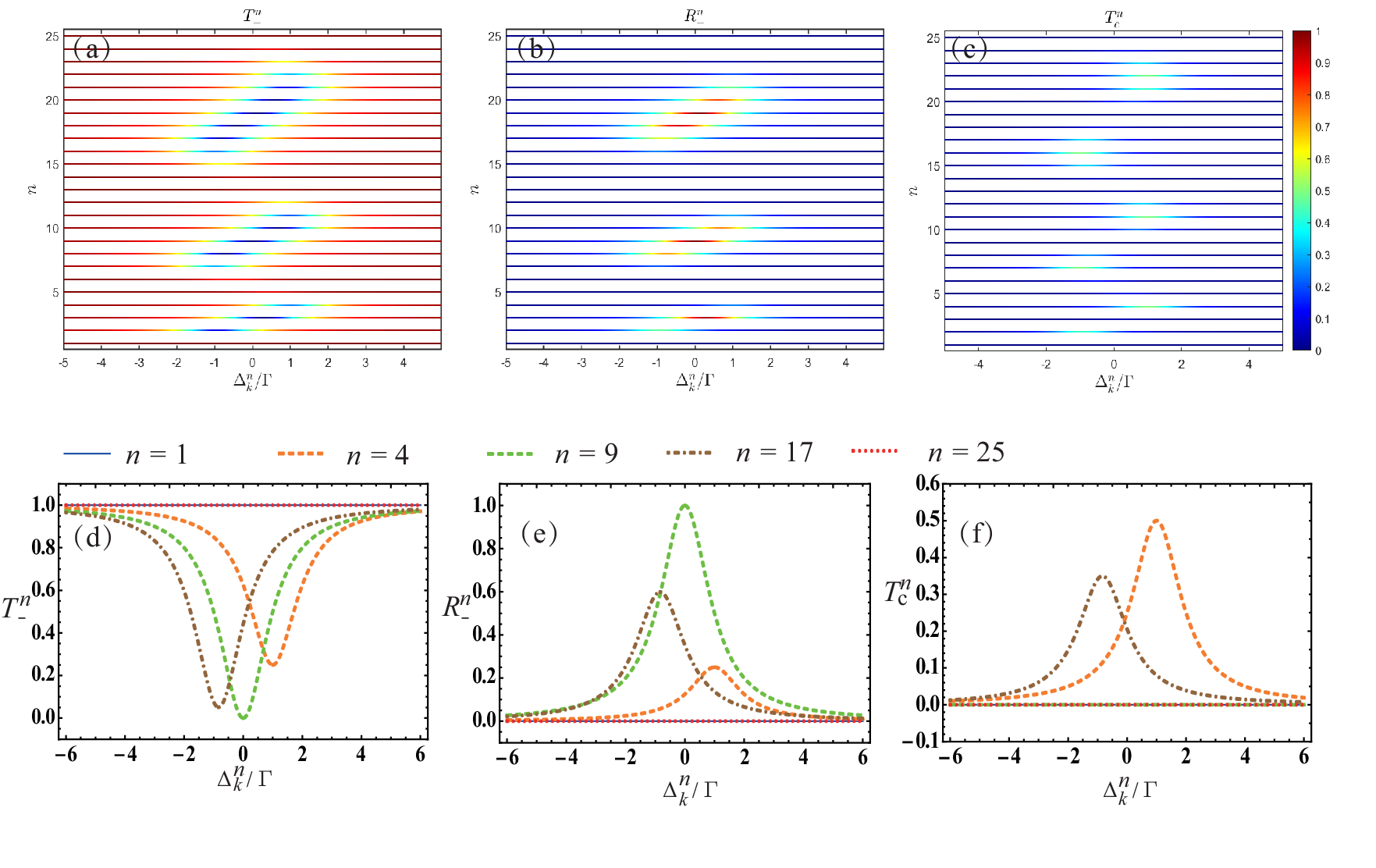}
\caption{The transmittance (a) $T_{-}^{n}$, reflectance (b) $R_{-}^{n}$, and
conversion probability (c) $T_{c}^{n}$ versus photon number $n$ and the
scaled detuning $\Delta _{k}^{n}/\Gamma $ when $|J_{1}|=|J_{2}|$, $\protect%
\omega _{fc}=0$, $\protect\omega _{e}/\Gamma =1015$, $g/\Gamma =15$, $%
\protect\varphi _{J}=0$, $\protect\tau \Gamma =0.1\protect\pi $. (d-f) The
profile of the scattering spectra at $n=1,4,9,17,25$.}
\label{fig3}
\end{figure*}


\section{\label{Sec:3}the scattering spectra and the conversion contrast}


The GA-resonator system forms an effective two-legged GA with multiple
levels in the presence of photons in the resonator. In the $\left(
n+1\right) $-excitation subspace with only one photon in the waveguide, the
GA-resonator system form an effective $\Lambda $-type GA, however, the
couplings of the GA to the left-going and right-going modes are equal at
each connection points, which can be explicitly found in Hamiltonian (\ref%
{Eq2-01}). The scattering spectra are determined by the following
parameters: the characteristic frequencies $\omega _{e}-\lambda _{n\pm }$,
the delay time $\tau $, the decay rate $\Gamma $, and the number $n$ of
photons in the resonator. Now, we will set $\left\vert J_{j}\right\vert =J$
since the system reduces to a two-level GA interaction with a 1D waveguide
and photonic bound states in continuum (BIC)\cite%
{GAB2PRR,GAB106PRL,GAB107PRA,GAB108PRA1,GAB109PRA,GAB6PRR,GAB109PRA1} can be
found for an injected photon when $n=0$, then $\gamma _{\pm }^{n}=\Gamma
_{\pm }^{n}\cos \varphi _{L}$. In Fig.~\ref{fig1}, we plot $\Gamma _{\pm
}^{n}$ and $\phi _{\pm }^{n}$ as a function of the photon number $n$ when $%
\omega _{fc}=0$, $\omega _{fc}\sim g$ and $\omega _{fc}\gg g$, respectively.
It can be observed from Fig.\ref{fig1}a that $\Gamma _{\pm }^{n}=\Gamma /2$
for resonance $\omega _{fc}=0$ (see the red dots and circles), $\Gamma
_{-}^{n}\approx \Gamma $ and $\Gamma _{+}^{n}\approx 0$ for lager detuning $%
\omega _{fc}\gg g$ (see the black dots and circles), and $\Gamma _{-}^{n}$
decreases but $\Gamma _{+}^{n}$ increases as $n$ increases for $\omega
_{fc}\sim g$ (see the blue dots and circles). Fig.~\ref{fig1}b shows that $%
\phi _{\pm }^{n}$ hardly varies with $n$ at lager detuning $\omega _{fc}\gg
g $ (see the black stars and diamonds), they all changes with $n$ at the
other two cases. In the following discussion, we will assume that the
resonator is on resonance with the $\left\vert g\right\rangle
\leftrightarrow \left\vert f\right\rangle $ transition of the GA, the
spontaneous damping rates $\Gamma _{\pm }^{n}$ and the nonlocal damping
rates $\gamma _{\pm }^{n}$ can be rewritten as $\Gamma /2$ and $\gamma _{\pm
}^{n}=\left( \Gamma /2\right) \cos \varphi _{J}$ as $\theta _{n}=\pi /2$ in Eq.(%
\ref{Eq2-02}) is fixed, however, $\phi _{\pm }^{n}$ takes different values
for different $n$. In a GA-waveguide system, $\tau ^{-1},\Gamma <\omega
_{e}-\lambda _{n\pm }$ is usually satisfied. Thus, the spectra of single
photons are decided by the propagation time $\tau $ and the lifetime $\Gamma
^{-1}$.

\subsection{Markovian regime}

In the Markovian regime, the delay time $\tau $ due to photon travelling
between coupling points is much smaller than the lifetime $\Gamma ^{-1}$.
the factors $\exp \left( {\pm }\mathrm{i}\Delta _{k}^{n}\tau \right) \approx
1$, Then the corresponding scattering amplitudes read
\begin{subequations}
\label{Eq3-01}
\begin{eqnarray}
t_{-}^{n} &=&\frac{\Delta _{k}^{n}-2\frac{J_{1-}^{n\ast }J_{2-}^{n}}{v}\sin
\phi _{-}^{n}+\mathrm{i}\left( \Gamma _{+}^{n}+\gamma _{+}^{n}e^{\mathrm{i}%
\phi _{+}^{n}}\right) }{\Delta _{k}^{n}+\mathrm{i}\left( \Gamma
_{+}^{n}+\gamma _{+}^{n}e^{\mathrm{i}\phi _{+}^{n}}\right) +\mathrm{i}\left(
\Gamma _{-}^{n}+\gamma _{-}^{n}e^{\mathrm{i}\phi _{-}^{n}}\right) } \\
r_{-}^{n} &=&-\frac{\mathrm{i}\gamma _{-}+\frac{\mathrm{i}}{v}\left(
\left\vert J_{1-}^{n}\right\vert ^{2}e^{-\mathrm{i}\phi _{-}^{n}}+\left\vert
J_{2-}^{n}\right\vert ^{2}e^{\mathrm{i}\phi _{n-}}\right) }{\Delta _{k}^{n}+%
\mathrm{i}\left( \Gamma _{+}^{n}+\gamma _{+}^{n}e^{\mathrm{i}\phi
_{+}^{n}}\right) +\mathrm{i}\left( \Gamma _{-}^{n}+\gamma _{-}^{n}e^{\mathrm{%
i}\phi _{-}^{n}}\right) } \\
r_{+}^{n} &=&\frac{\frac{\mathrm{i}e^{-\mathrm{i}\phi _{n}}}{v}\left(
J_{1+}^{n}e^{-\mathrm{i}\phi _{n+}}+J_{2+}^{n}\right) \left( J_{1-}^{n\ast
}+J_{2-}^{n\ast }e^{\mathrm{i}\phi _{-}^{n}}\right) }{\Delta _{k}^{n}+%
\mathrm{i}\left( \Gamma _{+}^{n}+\gamma _{+}^{n}e^{\mathrm{i}\phi
_{+}^{n}}\right) +\mathrm{i}\left( \Gamma _{-}^{n}+\gamma _{-}^{n}e^{\mathrm{%
i}\phi _{-}^{n}}\right) } \\
t_{+}^{n} &=&\frac{\frac{\mathrm{i}e^{-\mathrm{i}\phi _{n}}}{v}\left(
J_{1+}^{n}+J_{2+}^{n}e^{-\mathrm{i}\phi _{+}^{n}}\right) \left(
J_{1-}^{n\ast }+J_{2-}^{n\ast }e^{\mathrm{i}\phi _{-}^{n}}\right) }{\Delta
_{k}^{n}+\mathrm{i}\left( \Gamma _{+}^{n}+\gamma _{+}^{n}e^{\mathrm{i}\phi
_{+}^{n}}\right) +\mathrm{i}\left( \Gamma _{-}^{n}+\gamma _{-}^{n}e^{\mathrm{%
i}\phi _{-}^{n}}\right) }
\end{eqnarray}%
for light injected from the left side of the GA and
\end{subequations}
\begin{subequations}
\label{Eq3-02}
\begin{eqnarray}
\tilde{t}_{-}^{n} &=&\frac{\Delta _{k}^{n}-2\frac{J_{1-}^{n}J_{2-}^{n\ast }}{%
v}\sin \phi _{-}^{n}+\mathrm{i}\Gamma _{+}^{n}+\mathrm{i}\gamma _{+}^{n}e^{%
\mathrm{i}\phi _{+}^{n}}}{\Delta _{k}^{n}+\mathrm{i}\left( \Gamma
_{+}^{n}+\gamma _{+}^{n}e^{\mathrm{i}\phi _{+}^{n}}\right) +\mathrm{i}\left(
\Gamma _{-}^{n}+\gamma _{-}^{n}e^{\mathrm{i}\phi _{-}^{n}}\right) } \\
\tilde{r}_{-}^{n} &=&-\frac{\mathrm{i}\gamma _{-}+\frac{\mathrm{i}}{v}\left(
\left\vert J_{1-}^{n}\right\vert ^{2}e^{\mathrm{i}\phi _{-}^{n}}+\left\vert
J_{2-}^{n}\right\vert ^{2}e^{-\mathrm{i}\phi _{-}^{n}}\right) }{\Delta
_{k}^{n}+\mathrm{i}\left( \Gamma _{+}^{n}+\gamma _{+}^{n}e^{\mathrm{i}\phi
_{+}^{n}}\right) +\mathrm{i}\left( \Gamma _{-}^{n}+\gamma _{-}^{n}e^{\mathrm{%
i}\phi _{-}^{n}}\right) } \\
\tilde{r}_{+}^{n} &=&\frac{\frac{\mathrm{i}}{v}e^{\mathrm{i}\phi _{n}}\left(
J_{1+}^{n}e^{\mathrm{i}\phi _{+}^{n}}+J_{2+}^{n}\right) \left( J_{1-}^{n\ast
}+J_{2-}^{n\ast }e^{-\mathrm{i}\phi _{-}^{n}}\right) }{\Delta _{k}^{n}+%
\mathrm{i}\left( \Gamma _{+}^{n}+\gamma _{+}^{n}e^{\mathrm{i}\phi
_{+}^{n}}\right) +\mathrm{i}\left( \Gamma _{-}^{n}+\gamma _{-}^{n}e^{\mathrm{%
i}\phi _{-}^{n}}\right) } \\
\tilde{t}_{+}^{n} &=&\frac{\frac{\mathrm{i}}{v}e^{\mathrm{i}\phi _{n}}\left(
J_{1+}^{n}+J_{2+}^{n}e^{\mathrm{i}\phi _{+}^{n}}\right) \left( J_{1-}^{n\ast
}+J_{2-}^{n\ast }e^{-\mathrm{i}\phi _{-}^{n}}\right) }{\Delta _{k}^{n}+%
\mathrm{i}\left( \Gamma _{+}^{n}+\gamma _{+}^{n}e^{\mathrm{i}\phi
_{+}^{n}}\right) +\mathrm{i}\left( \Gamma _{-}^{n}+\gamma _{-}^{n}e^{\mathrm{%
i}\phi _{-}^{n}}\right) }
\end{eqnarray}%
for light injected from the right side of the GA. The Lamb shift $\delta
^{n}=\gamma _{+}^{n}\sin \phi _{+}^{n}+\gamma _{-}^{n}\sin \phi _{-}^{n}$
and the effective decay rate $\tilde{\Gamma}^{n}=\sum_{\alpha =\pm }\left(
\Gamma _{\alpha }^{n}+\gamma _{\alpha }^{n}\cos \phi _{\alpha }^{n}\right) $
are independent on the incident energy of the photon. In Fig.~\ref{fig2}, we
have plotted the transmittance and the conversion probability as a function
of $\Delta _{k}^{n}$ and a phase when $n=1$. All scattering coefficients
change periodically with the phase. The transmission spectra $T_{-}^{n}=%
\tilde{T}_{-}^{n}$ possess an anti-Lorentzian lineshape centered at $\delta
^{n}$ with a nonzero/zero minimum in Fig.~\ref{fig2}(a)/(c), the conversion
probability $T_{c}^{n}=\tilde{T}_{c}^{n}$ have the standard Lorentzian
lineshapes centered at $\delta ^{n}$ with a maximum $0.5$. The width of the
transmission spectra first increase and then decrease as $\phi _{\pm }^{n}$
changes from $-\pi $ to $\pi $ and reaches its maximum at $\phi _{\pm
}^{n}=0 $, so does the width of the conversion probability. One can also
observed a frequency-independent perfect transmission at $\varphi _{J}=2m\pi
$ and $\phi _{-}^{n}=\left( 2m+1\right) \pi $ in Fig.~\ref{fig2}(a) and a
transmittance dropping to zero at $\varphi _{J}=2m\pi $ and $\phi
_{+}^{n}=\left( 2m+1\right) \pi $ in Fig.~\ref{fig2}(c), however, the
frequency conversion is completely suppressed at either $\phi
_{-}^{n}=\left( 2m+1\right) \pi $ or $\phi _{+}^{n}=\left( 2m+1\right) \pi $
(see the red dotted line in \ref{fig2}b, \ref{fig2}d), which indicates that
reflectance $R_{-}^{n}=\tilde{R}_{-}^{n}=0$ at $\varphi _{J}=2m\pi $ and $%
\phi _{-}^{n}=\left( 2m+1\right) \pi $, and $R_{-}^{n}=\tilde{R}_{-}^{n}=1$
at $\delta ^{n}$ when $\varphi _{J}=2m\pi $ and $\phi _{+}^{n}=\left(
2m+1\right) \pi $ since the probability for the photon is conserved. The
frequency-independent perfect transmission can also be observed in Fig.~\ref%
{fig2}(e) when $\varphi _{J}=\left( 2m+1\right) \pi $ and $\phi
_{-}^{n}=2m\pi $. Substituting the parameters for the frequency conversion
vanishing into Eq.(\ref{Eq3-01}) and (\ref{Eq3-02}), we found that the
emission of the GA to the continuum of the $n_{-}$-channel ($n_{+}$-channel)
is completely suppressed at $\varphi _{J}=m\pi $ and $\phi _{-}^{n}=\left(
m+1\right) \pi $ for arbitrary $\phi _{+}^{n}$ ($\varphi _{J}=m\pi $ and $%
\phi _{+}^{n}=\left( m+1\right) \pi $ for arbitrary $\phi _{-}^{n}$), which
leads to the formation of GA-photon bound state\cite%
{GAB111PRA,GAB111PRA1,He25AQT8} in the $n_{-}$-channel ($n_{+}$-channel).
The GA-photon bound state gives rise to the BIC. Consequently, the number of
zeros in the frequency conversion is greater in Fig.~\ref{fig2}(f) than in
Fig.~\ref{fig2}(b,d). We note that the optimal $T_{c}^{n}$ is at most,
one-half when $\varphi _{J}=(m+1/2)\pi $ in Fig.~\ref{fig2}(f), $%
T_{c}^{n}\leq 1/2$ can also be observed in Fig.~\ref{fig2}(b,d) when $%
\varphi _{J}=m\pi $, this is because photon scattering coefficients with
either $\varphi _{J}=m\pi $ or $\phi _{-}^{n}=m\pi $ are symmetric for the
forward and backward propagating photons, which can be found from Eq.~(\ref%
{Eq3-01}a) and (\ref{Eq3-02}a).
\begin{figure}[tbp]
\includegraphics[width=8 cm,clip]{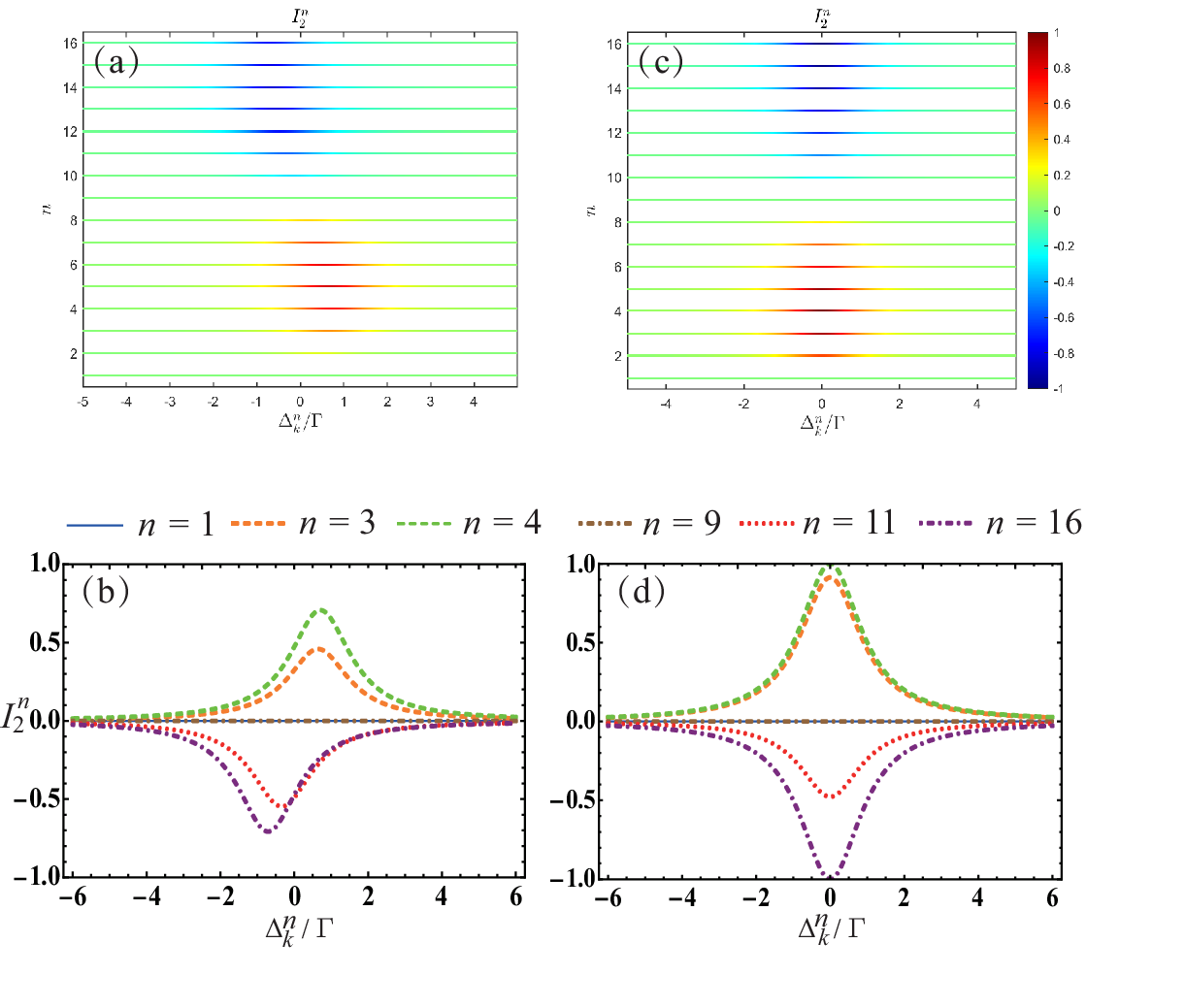} 
\caption{The conversion contrast $I_{2}^{n}$ versus photon number $n$ and the
scaled detuning $\Delta _{k}^{n}/\Gamma $ when $|J_{1}|=|J_{2}|$, $\protect%
\omega _{fc}=0$, $\protect\omega _{e}/\Gamma =1015$, $g/\Gamma =5$, $\protect%
\tau \Gamma =0.1\protect\pi $. (a) $\protect\varphi _{J}=0.25\protect\pi $,
(c) $\protect\varphi _{J}=0.5\protect\pi $. (b,d) The profile of the
conversion contrast at $n=1,3,4,9,11,16$.}
\label{fig4}
\end{figure}

To investigate the influence of the number $n$ of photons in the resonator
on the scattering spectra, we plot the transmittance, the reflectance and
the conversion probability versus the detuning and the photon number $n$ in
the resonator for $\varphi _{J}=0$ in Fig.\ref{fig3} with other parameters
satisfying a frequency-independent perfect transmission in the $1+1$
subspace. The scattering spectra are also periodic functions of $n$. As $n$
increases, the transmittance gradually decreases to zero at $\delta ^{n}$
then slowly increases back to 1. The transmittance reaches its minimum value
$T_{-}^{n}=0$ when $n=9$ while the reflectance reaches its maximum value $%
R_{-}^{n}=1$ at $\delta ^{n}$. Further increasing the photon number to $n=25$%
, frequency-independent perfect transmission is achieved again. However, the
conversion probabilities remains zero for any $\Delta _{k}^{n}$ in $1+1$, $%
9+1$ and $25+1$ subspace. These discussions demonstrate that under given
parameters with $\Gamma ,\tau $ etc. BICs can simultaneously emerge in the $%
n_{-}$-channel of different subspaces, which is evidenced by the perfect
transmission observed in both the $1+1$ and the $25+1$ subspaces.
Alternatively, BICs can occur in the $n_{-}$-channel of one subspace while
appearing in the $n_{+}$-channel of another subspace evidenced by the
perfect transmission in the $1+1$ subspace while perfect reflection in the $%
9+1$ subspace. In addition, BICs can also simultaneously emerge in the $%
n_{+} $-channel of different subspaces. All of these can be controlled by
setting the initial state of the resonator. The vanishing of frequency
conversion signifies the emergence of BIC. Thus, preventing the formation of
BICs enhances the conversion probability. However, the optimal conversion
probability is at most, one-half (see the maximum values in Fig.\ref{fig2}
and Fig.\ref{fig3}), which originates from the reciprocity of photon
transmission. In Fig.\ref{fig4}, we plot the influence of the
number $n$ of photons in the resonator on the conversion contrast and the
corresponding profiles at some subspaces with parameters $\omega _{e},g,\tau
,\Gamma $ presenting reciprocal spectra in the $1+1$ subspace. It can be
seen that $I_{2}^{n}=0$ when $n=1$, the conversion contrast first rises from $0$
to a maximum at $n=4$, later on returns to its original value $0$ at $n=9$,
then gradually drops to a minimum at $n=16$. This demonstrates that altering
the photon number in the resonator induces the non-reciprocal transmission
of single photons in the waveguide. The maximum and minimum values appear at
$\Delta _{k}^{n}=\delta ^{n}$, and $I_{2}^{n}$ achieves the optimal maximum $%
1$ and the optimal minimum $-1$ at position $\Delta _{k}^{n}=0$ when $%
\varphi _{J}=\pi /2$ (see Fig.~\ref{fig4}(d)).
\begin{figure}[tbp]
\includegraphics[width=8 cm,clip]{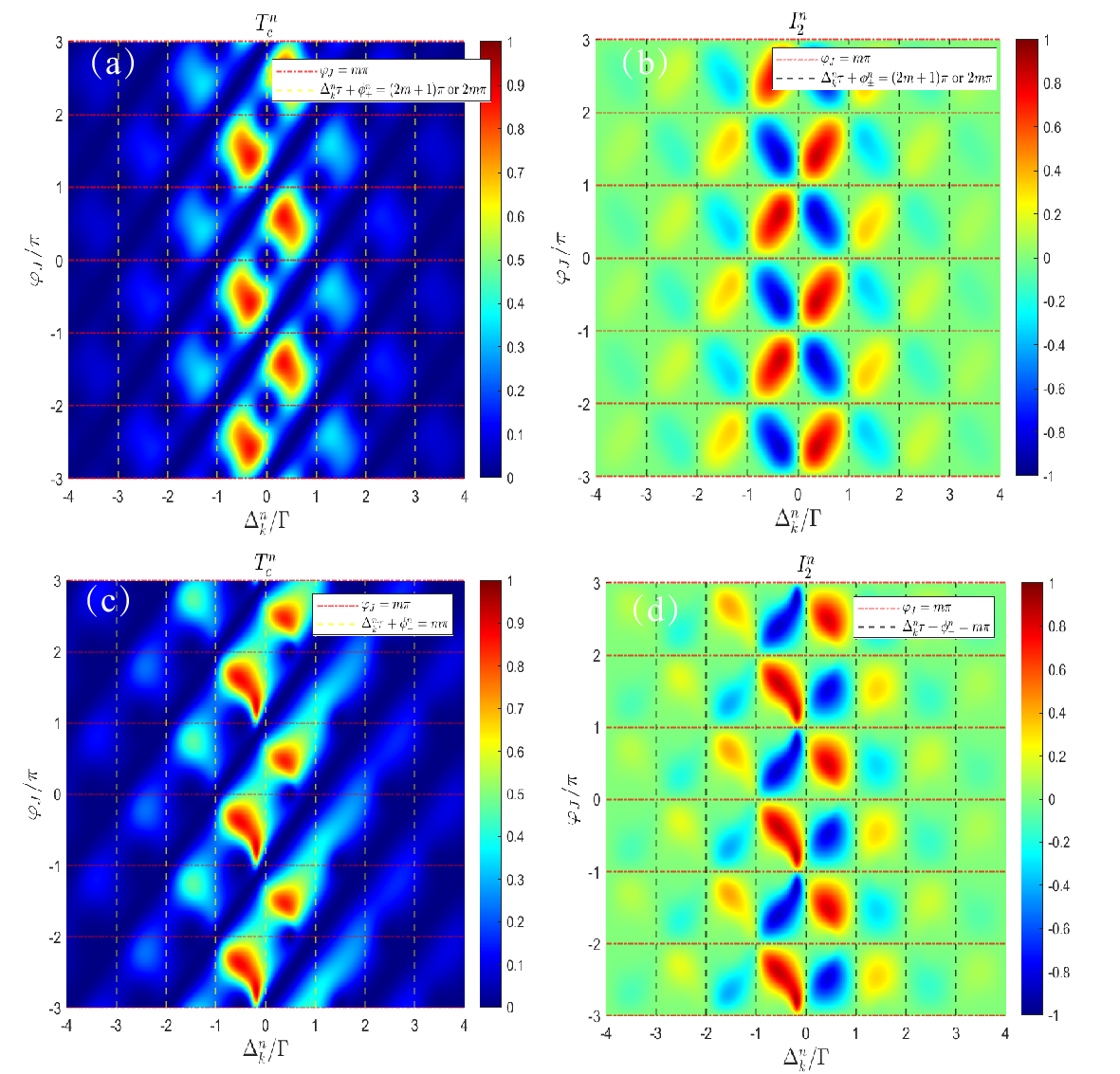} 
\caption{(a,c) The conversion probability $T_{c}^{n}$ and (b,d) the
conversion contrast $I_{2}^{n}$ versus phase $\protect\varphi _{J}$ and the
scaled detuning $\Delta _{k}^{n}/\Gamma $ for (a,b) $\protect\phi _{+}^{n}=%
\protect\pi $, (c,d) $\protect\phi _{+}^{n}=\protect\pi /2$, when $%
|J_{1}|=|J_{2}|$, $\protect\omega _{fc}=0$, $\protect\tau \Gamma =\protect%
\pi $, $\protect\phi _{-}^{n}=0$.}
\label{fig5}
\end{figure}

\subsection{Non-Markovian regime}

The Markovian approximation can be violated when the coupling between the
system and its environment is strong or when there is not a clear separation
between the typical timescales associated with the system and the
environment, here, we consider the situation with the travel time $\tau $ of
light between coupling points comparable to the characteristic timescale $%
\Gamma ^{-1}$. Phase factors are now detuning-dependent. In Fig.~\ref{fig5},
we have plotted the conversion probability $T_{c}^{n}$ and the conversion
contrast $I_{2}^{n}$ versus phase $\varphi _{J}$ and the scaled detuning $%
\Delta _{k}^{n}/\Gamma $ for $n=1$. When $\varphi _{J}=(m+1/2)\pi $,
different from the Lorentzian lineshape in the Markovian regime, the
conversion probability exhibits multiple peaks and dips. When $\varphi
_{J}=m\pi $, the conversion probability vanishes only at certain frequencies
with $\Delta _{k}^{n}\tau =m\pi $, see the intersection between the red
dotted line and the white dashed line in Fig.~\ref{fig5}a. We still can
observe that $T_{c}^{n}=0$ at locations $\Delta _{k}^{n}\tau =2m\pi $ when $%
\varphi _{J}=(2m+1)\pi $ in Fig.~\ref{fig5}c, however, the locations $\Delta
_{k}^{n}\tau =(2m+1)\pi $ of $T_{c}^{n}=0$ in Fig.~\ref{fig5}a are changed
to $\Delta _{k}^{n}\tau =(2m-1/2)\pi $ in Fig.~\ref{fig5}c. For $\varphi
_{J}=2m\pi $, $\Delta _{k}^{n}\tau =(2m+1)\pi $ also lead to $T_{c}^{n}=0$
in Fig.~\ref{fig5}c, however, locations $\Delta _{k}^{n}\tau =2m\pi $ in
Fig.~\ref{fig5}a are changed to $\Delta _{k}^{n}\tau =(2m+1/2)\pi $. The
locations $\Delta _{k}^{n}\tau =2m\pi $ ($\Delta _{k}^{n}\tau =(2m+1)\pi $)
of the conversion dips can be regarded as static zeroes for $\varphi
_{J}=(2m+1)\pi $ ($\varphi _{J}=2m\pi $). These static zeros are related to
phase $\phi _{-}^{n}$ since it remains constant in Fig.~\ref{fig5}a and \ref%
{fig5}c, the moving zeros are related to phase $\phi _{+}^{n}$, which help
us to find that the static zeros satisfy $\Delta _{k}^{n}\tau +\phi
_{-}^{n}=2m\pi $ for $\varphi _{J}=(2m+1)\pi $ or $\Delta _{k}^{n}\tau +\phi
_{-}^{n}=(2m+1)\pi $ for $\varphi _{J}=2m\pi $, resulting in complete
suppression of the emission from the $|e\rangle \leftrightarrow
|n_{-}\rangle $ transition to the waveguide; the moving zeros satisfy $%
\Delta _{k}^{n}\tau +\phi _{+}^{n}=2m\pi $ for $\varphi _{J}=(2m+1)\pi $ or $%
\Delta _{k}^{n}\tau +\phi _{+}^{n}=(2m+1)\pi $ for $\varphi _{J}=2m\pi $,
resulting in complete suppression of the emission from the $|e\rangle
\leftrightarrow |n_{+}\rangle $ transition to the waveguide. It can be
concluded that a zero conversion probability, in both Markovian and
non-Markovian regions, serves as a signature for the complete suppression of
the emission from the excited state to either of the lower states. In
addition, the conversion dips located at $\Delta _{k}^{n}\tau =2m\pi $ moves
toward the blue-shift direction as $\varphi _{J}$ increases in both Fig.~\ref%
{fig5}a and Fig.~\ref{fig5}c. We also found that the conversion dips occur
along the line $\Delta _{k}^{n}\tau +\phi _{-}^{n}-\varphi _{J}=(2m+1)\pi $
and $T_{c}^{n}\leq 1/2$ along the line $\varphi _{J}=m\pi $. The conversion
contrast always vanishes along the line $\varphi _{J}=m\pi $ since the
system is reciprocal for photons (see the red dotted line in Fig.~\ref{fig5}%
b and \ref{fig5}d), which is same as the conversion contrast in the
Markovain regime. However, a significant difference appear for both regimes
for $\varphi _{J}\neq m\pi $. It can be seen that the conversion probability
is always reciprocal leading to $I_{2}^{n}=0$ as long as $\phi _{-}^{n}=m\pi
$ in the Markovian regime, however, $I_{2}^{n}=0$ only occurs at certain
frequencies Fig.~\ref{fig5}(b,d), indicating that the system generally
exhibits non-reciprocity. This non-reciprocity is induced by non-Markovian
effects. It is easy to find from Fig.~\ref{fig5}b and \ref{fig5}d that $%
I_{2}^{n}=0$ as long as $\Delta _{k}^{n}\tau +\phi _{-}^{n}=m\pi $. Since $%
I_{2}^{n}=-I_{1}^{n}$, the disappearance of contrasts can be understood from
the interference of the transmitted wave. For a plane wave incident from the
left (right) of the giant atom, there are two paths propagating to the right
(left) of the giant atom: in the first path, the wave transmits directly
along the waveguide to the position $x>d/2$ ($x<-d/2$), accumulating a phase
of $\Delta _{k}^{n}\tau $, in the second path, waves propagate through the
GA to $x>d/2$ ($x<-d/2$), accumulating a phase of $\Delta _{k}^{n}\tau +\phi
_{-}^{n}+\varphi _{J}$ ($\Delta _{k}^{n}\tau +\phi _{-}^{n}-\varphi _{J}$).
The waves transmitting directly through the waveguide cancel each other. The
difference between these two waves undergoing the second path can be
regarded as the superposition of two waves traveling in the same direction,
except that there is a phase difference of between them, i.e., one wave
accumulates a phase of $\Delta _{k}^{n}\tau +\phi _{-}^{n}+\varphi _{J}$,
the other wave accumulates a phase of $\pi +\Delta _{k}^{n}\tau +\phi
_{-}^{n}-\varphi _{J}$. These two waves interference destructively when $%
\Delta _{k}^{n}\tau +\phi _{-}^{n}=m\pi $ or $\varphi _{J}=m\pi $.
Furthermore, we can also observe that in Fig.~\ref{fig5}b, for any given
phase $\varphi _{J}$, $I_{2}^{n}$ is symmetric about the point $\Delta
_{k}^{n}=0$, whereas this is not the case in \ref{fig5}d. For any given
detuning $\Delta _{k}^{n}$, both Fig.\ref{fig5}b and \ref{fig5}d exhibit
symmetry about the point $\varphi _{J}=0$.

\begin{figure}[tbp]
\includegraphics[width=8 cm,clip]{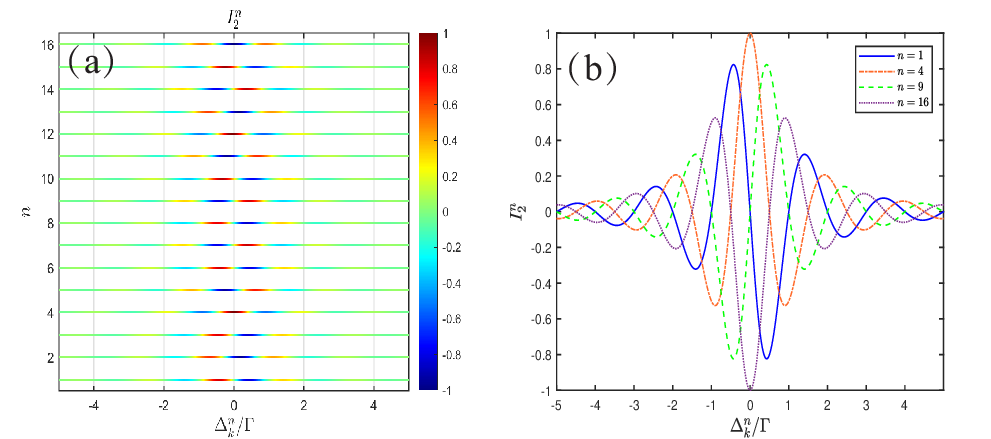} 
\caption{(a) The conversion contrast $I_{2}^{n}$ versus photon number $n$ and
the scaled detuning $\Delta _{k}^{n}/\Gamma $ when $|J_{1}|=|J_{2}|$, $%
\protect\omega _{fc}=0$, $\protect\omega _{e}/\Gamma =995.5$, $g/\Gamma =5.5$%
, $\protect\tau \Gamma =\protect\pi $, $\protect\varphi_{J}=0.5\protect\pi$.
(b) The profile of the conversion contrast at $n=1,4,9,16$.}
\label{fig6}
\end{figure}

To show that the number $n$ of photons in the resonator can tune the
transport properties of photons, we plot the conversion contrast versus the
scaling detuning and the photon number $n$ in the resonator when $\varphi
_{J}=\pi /2$ in Fig.\ref{fig6} with given parameters $\omega _{e},g,\tau
,\Gamma $. In the subspace with $n=1$, the maximum value of $I_{2}^{n}$
appears on the left side of $\Delta _{k}^{n}=0$, and the curve is symmetric
about the point $\Delta _{k}^{n}=0$ (see the blue solid line in Fig.\ref%
{fig6}b). As $n$ increases, the maximum initially shifts to the left, then
moves to the right. At $n=4$, $I_{2}^{n}$ reaches its maximum value of $1$,
which occurs exactly at $\Delta _{k}^{n}=0$, and the curve becomes symmetric
about the axis $\Delta _{k}^{n}=0$ (see the orange dash-dotted line in Fig.%
\ref{fig6}b). Subsequently, the value of $I_{2}^{n}$ decreases and moves to
the right, later oscillating on both sides of the axis $\Delta _{k}^{n}=0$.
Notably, the profile of $I_{2}^{n}$ at $n=9$ is the mirror image of that at $%
n=1$ with respect to the axis $\Delta _{k}^{n}=0$ (see the green dashed line
in Fig.\ref{fig6}b), and at $n=12$, the value of $I_{2}^{n}$ approaches
unity meaning $T_{c}^{n}\rightarrow 1$ and $\tilde{T}_{c}^{n}\rightarrow 0$,
and the location of this maximum lies close to $\Delta _{k}^{n}=0$. At $n=16$%
, a curve symmetric about the axis $\Delta _{k}^{n}=0$ reappears, but at
this time, $I_{2}^{n}$ attains its minimum value of $-1$ (see the purple
dotted line in Fig.\ref{fig6}b) indicating $T_{c}^{n}=0$ and $\tilde{T}%
_{c}^{n}=1$. One can see that the frequency conversion can be adjusted not
only by the accumulated phases but also by the initial state of the
resonator.

\section{Conclusion}

We have studied the transport of a single photon propagating in a 1D
conventional waveguide. The scattering target is a V-type GA. One of the
GA's transition is coupled to a 1D waveguide at two coupling points and the
other transition is coupled to a single-mode resonator. The presence of
photons in the resonator triggers the frequency conversion of single photons
in the 1D waveguide, and the frequency shift is controlled by the number of
photons in the resonator. The single-photon scattering spectra are obtained
by utilizing a real-space scattering method when the GA and the resonator
contain $n$ excitations. We analyze the influences of phase delay between
coupling points, the phase difference $\varphi_J$ between two GA-waveguide
couplings and the number of photons in the resonator on the scattering
spectra and the conversion contrast characterizing the non-reciprocity in
both the Markovian and the non-Markovian regimes. The spectra of the
conversion probability exhibit multiple peaks and dips in the non-Markovian
regime instead of the Lorentz lineshape in the Markovian regime. The zero
conversion probability, produced by the quantum interference induced by the
scale of a V-type GA characterized by phase delay $\phi^n_{\pm}$, serves as
a signature for the complete suppression of the emission from the excited
state to either of the lower states in both Markovian and non-Markovian
regions. The optimal maximum of the conversion probability is at most
one-half when the system is reciprocal. In particular, the conversion
contrast keep the constant value $0$ in the Markovian regime when $%
\varphi_J\neq m\pi$ and $\phi^n_{-}=m\pi$, one can break this reciprocity by
either changing the number of the excitations in the GA and the resonator in
the Markovian regime to enhance single-photon conversion efficiency to
unity, or increasing the distance of the two coupling points so that the
delay time is comparable to the lifetime of the GA. In the non-Markovian
regime, adjusting the number of the excitations in the GA and the resonator
changes the conversion contrast from the minimum to the maximum, which
achieves single-photon frequency up- or down-conversion with unity
efficiency.

\begin{acknowledgments}
This work was supported by NSFC Grants No.12421005, No.12247105, XJ-Lab Key Project (23XJ02001),
and the Science $\And $ Technology Department of Hunan Provincial Program (2023ZJ1010).
\end{acknowledgments}

\end{subequations}

\end{document}